# Reconstructing Faces from fMRI Patterns using Deep Generative Neural Networks

**Rufin VanRullen[1,*] and Leila Reddy[1]**

1.  CerCo, CNRS, UMR 5549, Université de Toulouse, Toulouse, 31052 (France)
*   Corresponding author : rufin.vanrullen@cnrs.fr

**Abstract:** While objects from different categories can be reliably decoded from fMRI brain response patterns, it has proved more difficult to distinguish visually similar inputs, such as different instances of the same category. Here, we apply a recently developed deep learning system to the reconstruction of face images from human fMRI patterns. We trained a variational auto-encoder (VAE) neural network using a GAN (Generative Adversarial Network) unsupervised training procedure over a large dataset of celebrity faces. The auto-encoder latent space provides a meaningful, topologically organized 1024-dimensional description of each image. We then presented several thousand face images to human subjects, and learned a simple linear mapping between the multi-voxel fMRI activation patterns and the 1024 latent dimensions. Finally, we applied this mapping to novel test images, turning the obtained fMRI patterns into VAE latent codes, and ultimately the codes into face reconstructions. Qualitative and quantitative evaluation of the reconstructions revealed robust pairwise decoding (>95% correct), and a strong improvement relative to a baseline model (PCA decomposition). Furthermore, this brain decoding model can readily be recycled to probe human face perception along many dimensions of interest; for example, the technique allowed for accurate gender classification, and even to decode which face was imagined, rather than seen by the subject. We hypothesize that the latent space of modern deep learning generative models could serve as a valid approximation for human brain representations.

Decoding sensory inputs from brain activity is both a modern technological challenge and a fundamental neuroscience enterprise. Multi-voxel fMRI pattern analysis, inspired by machine learning methods, has produced impressive "mind-reading" feats over the last 15 years[1-4]. A notoriously difficult problem, however, is to distinguish brain activity patterns evoked by visually similar inputs, such as objects from the same category, or distinct human faces[5-9]. Here, we propose to take advantage of recent developments in the field of deep learning. Specifically, we use a variational auto-encoder or VAE[10], trained with a Generative Adversarial Network (GAN) procedure[11,12], as illustrated in Figure 1A. The resulting VAE-GAN model is a state-of-the-art deep generative neural network for face representation, manipulation and reconstruction[12]. The "face latent space" of this network provides a description of numerous facial features that could approximate face representations in the human brain. In this latent space, faces and face features (e.g., maleness) can be represented as linear combinations of each other, and different concepts (e.g., male, smile) can be manipulated using simple linear operations (Figure 1B). The versatility of this deep generative neural network latent space suggests a possible homology with human brain facial representations, and makes it an ideal candidate for fMRI-based face decoding. We thus reasoned that it could prove advantageous, when decoding brain activity, to learn a mapping between the space of fMRI patterns and this kind of latent space, rather than the space of image pixels (or a linear combination of those pixels, as done in recent state-of-the-art approaches involving PCA[13,14]). In particular, we surmised that the VAE-GAN model captures and untangles most of the complexity of human face representations, flattening and evening up the "face manifold" as human brains might do[15], so that simple linear brain decoding methods can suffice. In line with this hypothesis, we find that the technique outperforms a current (non-deep learning) state-of-the-art method, and not only allows to reconstruct a reliable estimate of seen faces, but also to decode face gender or face mental imagery. In sum, our study's contributions are (at least) threefold:

1. We introduce a new, state-of-the-art brain decoding method, based on the latest developments in deep learning and generative models.
2. We propose that many outstanding questions about face processing in the human brain could be addressed using this method and our large-scale (publicly available) fMRI dataset. We illustrate this proposal with two examples, gender processing and mental imagery, in both cases with results that go beyond the previous state-of-the-art.
3. We speculate that the latent space of deep generative models may be homologous to human brain representations.

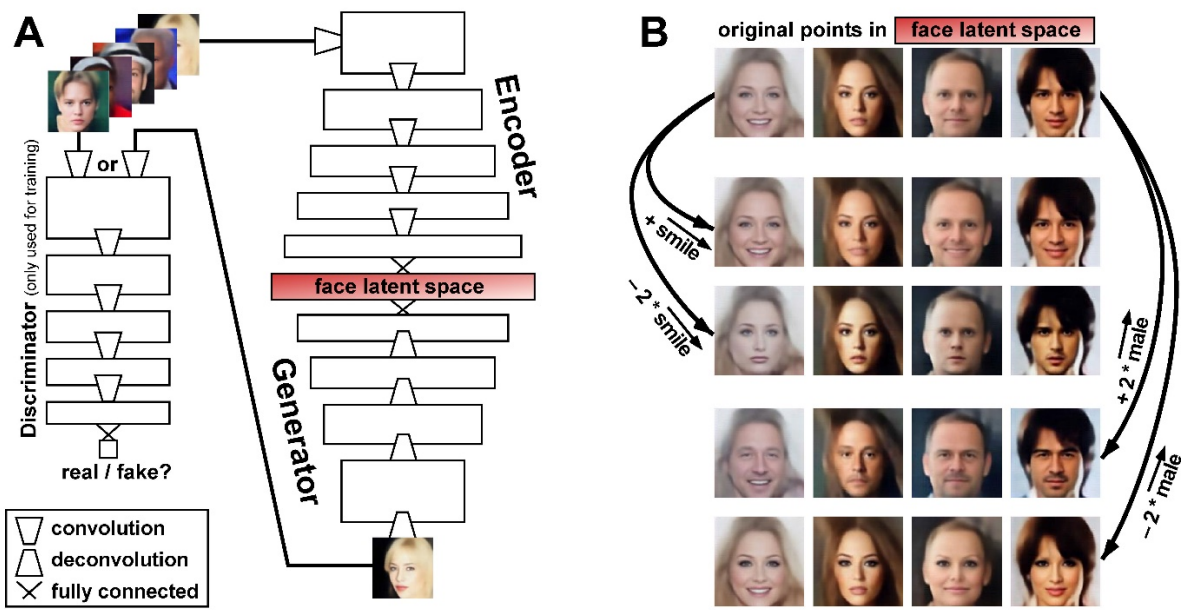

**Figure 1**. **Deep neural network latent space. A. VAE-GAN Network Architecture**. Three networks learn complementary tasks. The Encoder network maps a face image onto a latent representation (1024-dimensional), shown in red, which the Generator network converts into a novel face image. The Discriminator network (only used during the training phase) outputs a binary decision for each given image, either from the original dataset, or from the Generator output: is the image real or fake? Training is called "adversarial" because the Discriminator and Generator have opposite objective functions. (For simplicity, this diagram does not reflect the fact that the VAE latent space is actually a variational layer, which samples latent vectors stochastically from a probability distribution). **B. Latent space properties.** Once training is complete, the VAE latent space can be sampled and manipulated with simple linear arithmetic. The top row shows four original faces. The lower rows show the result of linear operations on the sample faces. For example, adding or subtracting a "smile vector" $\overrightarrow{smile}$ (computed by subtracting the average latent description of 1000 faces having a "no-smile" label from the average latent description of 1000 faces having a "smile" label) creates images of the original faces smiling or frowning (2nd and 3rd rows). The same operation can be done by adding or subtracting (a scaled version of) the average vector $\overrightarrow{male}$ (4th and 5th rows), making the original faces more masculine or more feminine. In short, the network manipulates face-related "concepts", which it can extract from and render to pixel-based representations.

# Results

## Face decoding and reconstruction

We used the pre-trained VAE-GAN model described in Figure 1 (with "frozen" parameters) to train a brain decoding system. During training (Figure 2A), the system learned the correspondence between brain activity patterns in response to numerous face images and the corresponding 1024-D latent representation of the same faces within the VAE network. More than 8,000 distinct examples were used on average (range across subjects: [7664-8626]), which involved 12 hours of scanning over 8 separate sessions for each subject. The learning procedure assumed that each brain voxel's activation could be described as a weighted sum of the 1024 latent parameters, and we simply estimated the corresponding weights via linear regression (GLM function in SPM; see Methods). After training (Figure 2B), we inverted the linear system, such that the decoder was given the brain pattern of the subject viewing a specific, novel face image as input (a face that was not included in the training set), and its output was an estimate of the 1024-dimensional latent feature vector for that face. The image of the face was then generated (or "reconstructed") through the generative (VAE-GAN) neural network.

We contrasted the results obtained from this deep neural network model with those produced by another, simpler model of face image decomposition: principal components analysis (PCA, retaining only the first 1024 principal components from the training dataset; see Supplementary Figure S1). The PCA model also describes every face by a vector in a 1024-dimensional latent space, and can also be used to reconstruct faces based on an estimate of this 1024-D feature vector, as demonstrated in recent studies[13,14].

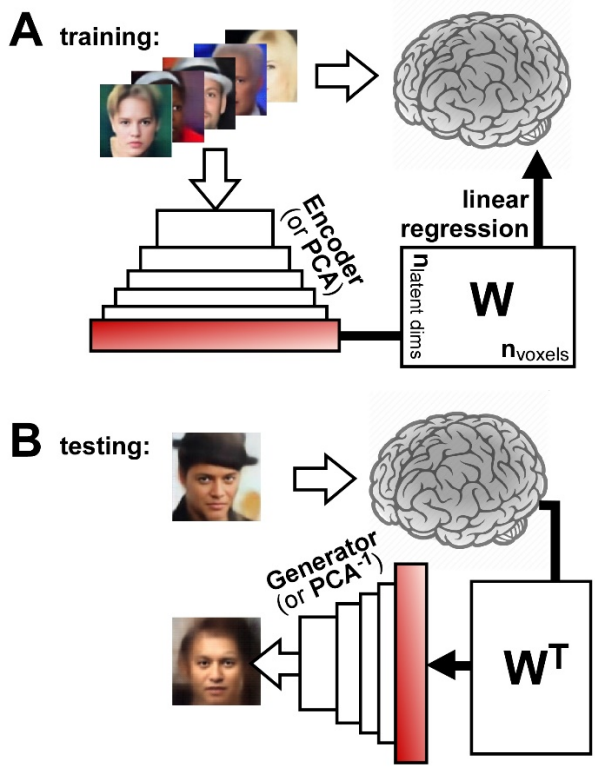

**Figure 2. Brain decoding of face images based on VAE-GAN latent representations. A. Training phase.** Each subject saw more than 8,000 faces (one presentation each) in a rapid event-related design. The same face images were also run through the "Encoder" network (as described in Figure 1) or a PCA decomposition, to obtain a 1024-dimensional latent face description. The "brain decoder" was a simple linear regression, trained to associate the 1024-dimensional latent vector with the corresponding brain response pattern. This linear regression, with 1024 parametric regressors for the BOLD signal (and an additional constant "bias" term), produced a weight matrix W (1025 by $n_{voxels}$ dimensions) optimized to predict brain patterns in response to face stimuli. **B. Testing phase.** We also presented 20 distinct "test" faces (not part of the training set; at least 45 randomly interleaved presentations each) to the subjects. The resulting brain activity patterns were simply multiplied by the transposed weight matrix $W^T$ ($n_{voxels}$ by 1025 dimensions) and its inverse covariance matrix to produce a linear estimate of the latent face dimensions. The Generator network (Figure 1A) or an inverse PCA transform was then applied to translate the predicted latent vector into a reconstructed face image.

For both the deep neural network and PCA-based models, we defined a subset of the gray matter voxels as our "region-of-interest". Indeed, many parts of the brain perform computations that are not related to face processing or recognition; entering such regions in our analysis would adversely affect signal-to-noise. Our selection criterion combined two factors: (i) voxels were expected to respond to face stimuli (as determined by a t-test between face and baseline conditions, i.e. fixation of an empty screen), and (ii) the explained variance of the voxels' BOLD response was expected to improve when the 1024 latent face features were entered as regressors in the linear model (compared to a baseline model with only a binary face regressor: face present/absent). The distribution of voxels along these two dimensions, and the corresponding selection criterion, are illustrated for one representative subject in Supplementary Figure S2. Across the four subjects, the number of resulting voxels in the selection was approximately 100,000 (mean: 106,612; range: [74,183-162,388]). The selected voxels are depicted in Figure 3; they include occipital, temporal, parietal and frontal regions. A separate selection was made based on the PCA face parameters, and used for the PCA-based "brain decoder" (mean number of selected voxels: 106,685; range: [74,073-164,524]); the selected regions were virtually identical for the two models (not shown). It is important to highlight that the above voxel selection criteria were applied based on BOLD responses to the training face images only, but not to the 20 test images; therefore, the decoding analysis does not suffer from "circular reasoning" issues caused by this voxel selection[16].

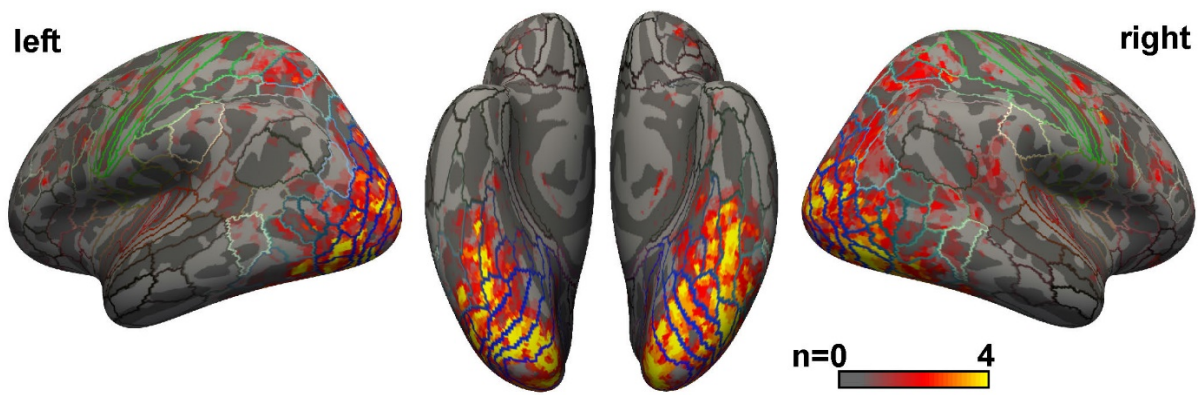

**Figure 3. Voxels selected for brain decoding.** Voxels were selected based on a combination of their visual responsiveness and their GLM goodness-of-fit during the brain decoder training stage (Figure 2A). The color code (red to yellow) indicates the number of subjects (1 to 4) for whom each particular voxel was selected. The colored lines indicate the boundaries of standard cortical regions[17].

Examples of the reconstructed face images from the test image set of each of the four subjects are shown in Figure 4A. While both the VAE-GAN and the PCA models could reconstruct an acceptable likeness of the original faces, the images reconstructed from the deep generative neural network (VAE-GAN) appear more realistic, and closer to the original image. We quantified the performance of our brain decoding system by correlating the brain-estimated latent vectors of the 20 test faces with the 20 actual vectors, and used the pairwise correlation values to measure the percentage of correct classification. For each subject, for each of the 20 test faces, we compared the decoded 1024-D vector to the ground-truth vector from the actual test image, and to that of another test image (distractor): brain decoding was "correct" if the correlation with the actual target vector was higher than with the distractor vector. This was repeated for all (20*19) pairs of test images, and the average performance compared to chance (50%) with a non-parametric Monte-Carlo test (see Methods: Statistics). Reconstructions from the GAN model achieved 95.5% classification (range: [91.3%-98.7%], all $p<10^{-6}$), while the PCA model only reached 87.5% (range [76.6%-92.4%], still highly above chance, all $p<10^{-4}$, but much below the GAN model, Friedman non-parametric test, $\chi^2(1)=4$, $p<0.05$). We also tested the ability of the brain decoder to pick the exact correct face among the 20 test faces: this "full recognition" task was deemed correct if and only if the reconstructed latent vector was more correlated to the true target vector than to *all* of the 19 distractor vectors. This is a more stringent test of face recognition, with chance level at 5%: the VAE-GAN model achieved 65% correct (range: [40%-75%], binomial test, all $p<10^{-6}$), while the PCA model resulted in 41.25% correct recognition only (range [25%-50%], all $p<10^{-3}$); again, the VAE-GAN model performance was significantly higher than the PCA ($\chi^2(1)=4$, $p<0.05$).

As linear regression models typically require many more data samples than their input dimensions, we had initially decided to train the brain decoding system with ~8,000 faces per subject (compared with the 1,024 latent dimensions). In order to establish whether smaller training sets might be sufficient, we repeated the linear regression step (computation of the W matrix in Figure 2A) using only one half, one quarter or one eighth of the training dataset (see Supplementary Figure S3). For both pairwise and full recognition measures, above-chance performance could already be obtained with ~1,000 training faces; however, decoding performance kept growing as the training set size was increased, and was highest for ~8,000 training faces. Importantly, the PCA model remained below the VAE-GAN model for all training set sizes.

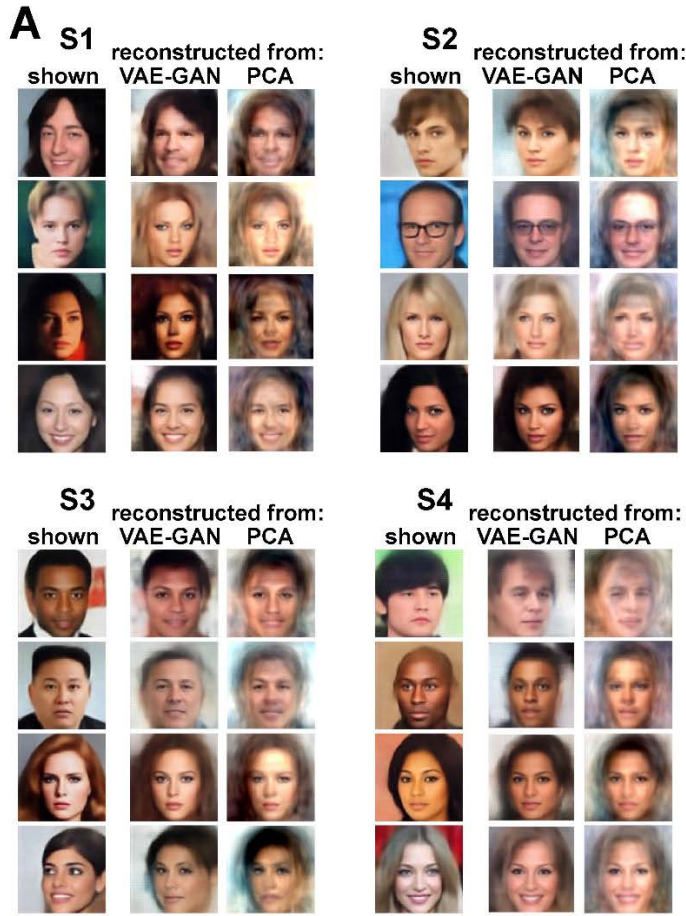

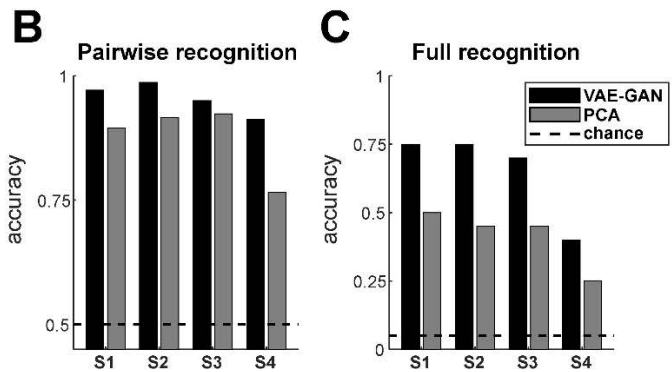

**Figure 4. Face reconstruction. A. Examples of reconstructed face images.** For each of our four subjects (S1-S4), the first column displays four example faces (two male + two female, chosen among the 20 test faces) actually shown to the subject during the scanning sessions. The next two columns are the face reconstructions based on the corresponding fMRI activation patterns for the brain decoding system trained using the VAE-GAN latent space (middle column) or PCA decomposition (right column). **B. Pairwise recognition.** The quality of brain decoding was quantified with a pairwise pattern classification (operating on the latent vector estimates), and the average performance compared to chance (50%). Brain decoding from the VAE-GAN model achieved 95.5% correct performance on average ($p<10^{-6}$), the PCA model only 87.5% ($p<10^{-4}$); the difference between the two models was significant ($\chi^2(1)=4$, $p<0.05$). **C. Full recognition.** A more stringent performance criterion was also applied, whereby decoding was considered correct if and only if the procedure identified the exact target face among all 20 test faces (chance=5%). Here again, performance of the VAE-GAN model (65%) was far above chance ($p<10^{-6}$), and outperformed ($\chi^2(1)=4$, $p<0.05$) the PCA model (41.25%; $p<10^{-3}$).

These comparisons indicate that it is easier and more efficient to create a linear mapping from human brain activations to the VAE-GAN latent space than to the PCA space. This is compatible with our hypothesis that the deep generative neural network is more similar to the space of human face representations. In addition, this classification accuracy was measured here based on the distance (or vector correlation) in the latent space of each model; it is even possible that the difference between the two models could be exacerbated if their accuracy was evaluated with a common metric, such as the perceptual quality of reconstructed images. To support this idea, we asked naïve human observers to compare the quality of faces reconstructed by the two models: each original test image from each of the four subjects was shown together with the corresponding VAE-GAN and PCA reconstructions; the observer decided which reconstruction was perceptually more similar to the original. Each pair was rated 15 times overall, by at least 10 distinct participants, with at least 5 participants seeing the two response options in either order, VAE-GAN first or PCA first. The VAE-GAN reconstruction was chosen in 76.1% of trials, while the PCA reconstruction only in 23.9% of trials. That is, observers were three times more likely to prefer the quality of VAE-GAN reconstructed faces than PCA reconstructions, a difference that was highly unlikely to occur by chance (binomial test, 1200 observations, $p<10^{-10}$).

**Contributions from distinct brain regions**

To determine which brain regions most contributed to the face reconstruction abilities of the two brain decoding models, for each subject we divided our voxel selection into three equally-sized subsets, as illustrated in Figure 5A. The brain decoding and face reconstruction procedure was then applied separately for these 3 subsets. The pairwise recognition results revealed that occipital voxels, and to a lesser extent temporal voxels, were providing most of the information necessary for brain decoding (Figure 5B). Occipital voxels decoding performance was much above chance (50%) for both models (VAE-GAN: 91.8%, all individual $p<10^{-6}$; PCA: 87.2%, all $p<10^{-4}$), and similarly for temporal voxels (VAE-GAN: 78.8%, all $p<10^{-3}$; PCA: 73.6%, all $p<0.01$). On the other hand, frontoparietal voxels, although they satisfied our selection criteria (see Figure 3) did not carry sufficiently reliable information on their own to allow for accurate classification (VAE-GAN: 60.1%, one subject with $p<10^{-6}$, all other $p>0.2$; PCA: 56.4%, one subject with $p<10^{-6}$, all other $p>0.05$; see, however, Lee et al[14]). The pattern of results was identical for both the VAE-GAN and the PCA-based decoding models: a non-parametric Friedman test suggested that performance differed across the three subsets (for VAE-GAN: $\chi^2(2)=8$, $p<0.02$; for PCA: $\chi^2(2)=6.5$, $p<0.04$), with post-hoc tests revealing that occipital voxels performed significantly better than frontoparietal ones, with temporal voxels in-between (not significantly different from either of the other two). Across all voxel selections, PCA always produced lower accuracies than VAE-GAN—though this difference did not reach statistical significance given our limited subject number (across all 3 voxel selections, $\chi^2(1)\geq3$, $p>0.08$).

To further distinguish the relative contributions of the three brain regions to the brain decoding performance, we also employed a variance partitioning approach (Figure S4). Compatible with the results already described in Figure 5B, we found that latent vector predictions derived from occipital voxels accounted for the largest portion of the variance of the corresponding ground-truth latent vectors, followed by temporal voxels, and finally frontoparietal voxels. Each of the three areas also had a unique, independent contribution to the explained variance, which was sizably larger for the VAE-GAN than the PCA model. That is, even though occipital voxels provided the most accurate reconstructions, temporal voxels did not merely convey redundant information.

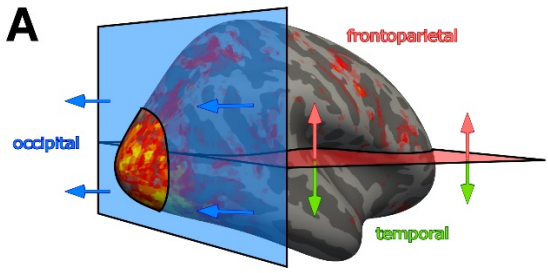

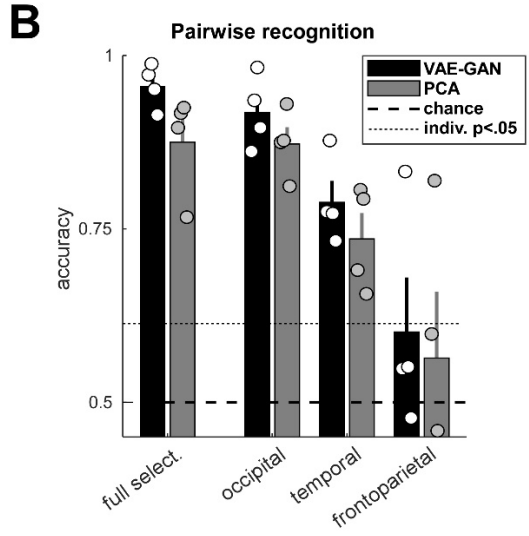

**Figure 5. Contributions from distinct brain regions. A. voxel segmentation procedure.** To investigate the brain regions that most strongly supported our brain decoding performance, while keeping the different subsets comparable, we linearly separated our voxel selection into 3 equally sized subsets. First, the 1/3 of most posterior voxels for each subject were labelled as "occipital". Among the remaining voxels, the more rostral half (1/3 of the initial number) was labelled as "temporal", and the remaining caudal half as "frontoparietal". This 3-way segmentation, different for each subject, was chosen because the performance of our brain decoding procedure is highly sensitive to the number of included voxels. **B. Pairwise recognition performance for the different regions of interest.** The full selection refers to the set of voxels depicted in Figure 3; it is the same data as in Figure 4B, averaged over subjects (error bars reflect standard error of the mean). Circles represent individual subjects' performance. The dotted line is the p<.05 significance threshold for individual subjects' performance. Among the 3 subsets, and for both the VAE-GAN and PCA models, performance is maximal in occipital voxels, followed by temporal voxels. Frontoparietal voxels by themselves do not support above-chance performance (except for one of the four subjects). In all cases, the VAE-GAN model performance remains higher than the PCA model.

**Possible applications: Gender decoding as example**

The learned mapping between the brain activation patterns and the deep generative neural network latent space (i.e., the matrix W in Figure 2A) can serve as a powerful tool to probe the human brain representation of faces, without necessarily having to perform costly additional experiments. A straightforward application, for example, could be the visualization of the facial feature selectivity of any voxel or ROI in the brain. The voxel or ROI defines a subset of columns in the W matrix (Figure 2), each column storing a latent vector that represents the voxel's facial selectivity. By simply running this latent vector (or its average over the ROI) into the face Generator network, the voxel or ROI selectivity can be revealed as an actual face image.

Another extension would be to explore the brain representation of behaviorally important facial features, such as gender, race, emotion or age. Any such face property can be expressed as a latent vector, which can easily be computed based on a number of labelled face examples (by subtracting the average latent vector for faces without the attribute label from the average latent vector for faces with the label; see Figure 1B for examples of latent vectors computed with faces having a "smile" label, or a "male" label). The publicly available celebrity face dataset (CelebA[18]) used in our experiments is already associated with 40 such labels describing gender, expressions, skin or hair color and numerous other properties of each face. Note that these 40 binary labels (feature present/absent) were collected via a manual annotation procedure for each face stimulus in the face dataset, and were chosen to be representative of the variability in the dataset. Given the latent vector describing such a facial property, we can use the brain decoding model to find out which brain voxels are most sensitive to the associated face property. This procedure is illustrated in Figure S5 for the example of the "gender" attribute ("male" label). The voxels most sensitive to this facial property are recovered by computing the column-wise correlation of the matrix W with the "male" latent vector: gender-selective voxels must have strongly positive or strongly negative correlation values (depending on their preference towards male or female faces). The voxels with largest (absolute-value) correlations are found in occipital and temporal regions, notably in both early visual areas and the fusiform cortex (Figure S5), consistent with a previous report of distributed representation of gender information[6].

Finally, another way to investigate the brain representation of a specific facial attribute is to create a simple classifier to label the brain-decoded latent vectors according to this face property. This is illustrated in Figure 6, again for the example of the "gender" face attribute. Each brain-decoded latent vector is projected onto the "gender" axis of the latent space (Figure 6A), and the sign of the projection determines the classification output ("male" for positive, "female" for negative signs). This rudimentary classifier provides sufficient information to classify face gender with 70% accuracy (binomial test, p=0.0001; Figure 6B). A non-parametric Friedman test indicates that gender decoding performance differs across the three subsets ($\chi^2(2)$=7.6, p<0.03), and a post-hoc test reveals that occipital voxels perform significantly better than frontoparietal ones, with temporal voxels in-between (not significantly different from either of the other two). Previous attempts at classifying face gender using multi-voxel pattern analysis had achieved limited success, with maximum classification accuracy below 60%[6,8]. Our simple linear brain decoder (Figure 6A) already improves on these previous methods, while still leaving room for future enhancements, e.g. using more powerful classification techniques (such as SVM) on the brain-decoded latent vectors.

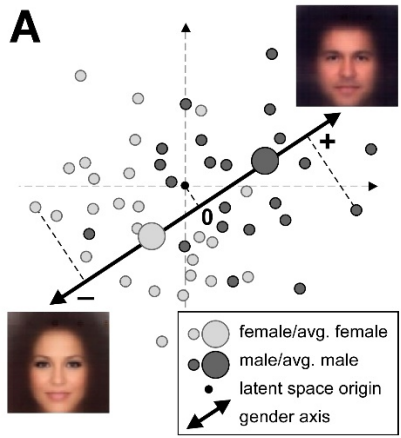

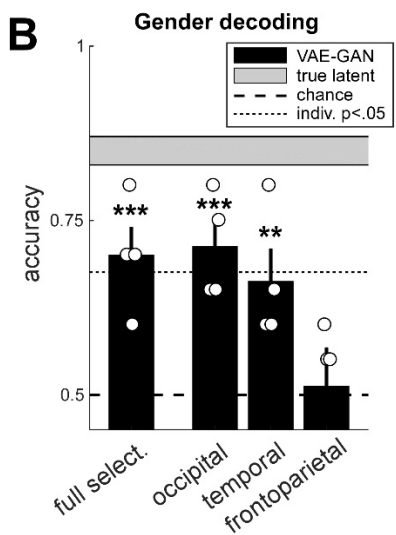

**Figure 6. Gender decoding. A. Basic linear classifier.** A simple gender classifier was implemented as a proof-of-principle. The "gender" axis was computed by subtracting the average latent description of 10,000 female faces from the average latent description of 10,000 male faces. Each latent vector was simply projected onto this "gender" axis, and positive projections were classified as male, negative projections as female. **B. Decoding accuracy.** When applied to the true latent vectors for each subject's test faces, this basic classifier performed at 85% correct (range: [80%-90%]). This is the classifier's ceiling performance, represented as a horizontal gray region (mean±sem across subjects). When operating on the latent vectors estimated via our brain decoding procedure, the same gender classifier performed at 70% correct, well above chance (binomial test, p=0.0001; bars represent group-average accuracy ±sem across subjects, circles represent individual subjects' performance). Gender classification was also accurate when restricting the analysis to occipital voxels (71.25%, p=0.00005) or temporal voxels (66.25%, p<0.001), but not frontoparietal voxels (51.25%, p=0.37). The star symbols indicate group-level significance: *** for p<0.001, ** for p<0.01. The dotted line is the p<.05 significance threshold for individual subjects' performance.

**Imagery decoding**

To further demonstrate the versatility of our brain decoding method, we next applied it to another notoriously difficult problem: retrieving information about stimuli that are not directly experienced by the subject, but only imagined in their "mind's eye". Previous studies have shown that this classification problem can be solved when the different classes of stimuli to be imagined are visually distinctive[19], such as images from different categories[20-25]. However, the ability to distinguish between highly visually similar objects—such as different faces—during imagery, as far as we know, has not been reported before.

Prior to the experiment, each subject chose one face among a set of 20 possible images (different from both their training and test image sets). During the experiment, they were instructed to imagine this specific face, whenever a large gray square occurred in the middle of the screen (12s presentation). These imagery trials were repeated 52 times on average (range across subjects: [51-55]) during the fMRI scanning sessions, interleaved with normal stimulus presentations. The average BOLD response during imagery was then used to estimate a latent face vector (using the brain decoder illustrated in Figure 2B), and this vector was compared to the 20 possible latent vectors in a pairwise manner, as described previously for test images (Figures 4B, 5B). The pairwise decoding performance was not different from chance (50%) in each of our predefined regions of interest (full selection p=0.53, occipital p=0.30 or frontoparietal regions p=0.43), with the sole exception of the temporal voxel selection, which produced 84.2% correct decoding (p=0.012). A non-parametric Friedman test indicated that imagery decoding performance differed across the three subsets ($\chi^2(2)=6.5$, p<0.04), and a post-hoc test revealed that temporal voxels performed significantly better than frontoparietal ones, with occipital voxels in-between (not significantly different from either of the other two). Altogether, temporal regions, but not occipital or frontoparietal ones, can support mental imagery reconstruction. This performance could reflect the strong involvement of temporal brain regions in high-level face processing[26-28], as well as the primarily top-down nature of mental imagery[29]. In any case, the ability to classify imagined faces from brain response patterns highlights again the flexibility and potential of our approach.

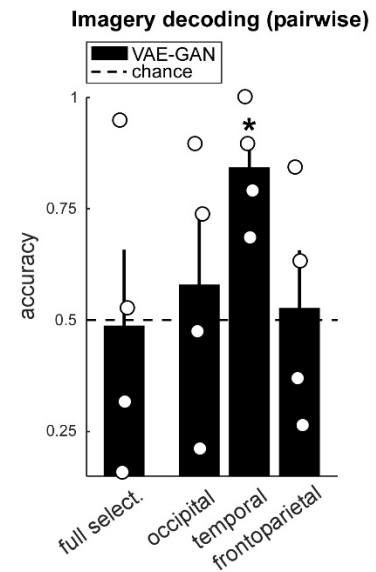

**Figure 7. Imagery decoding.** The fMRI BOLD response pattern recorded during mental imagery of a specific face (not visible on the screen) was passed through our brain decoding system. The resulting estimated latent vector was compared to the true vector and 19 distractor vectors, in a pairwise manner. Only the temporal voxel selection supported above-chance imagery decoding, with 84.2% correct performance (p=0.012). Neither occipital, nor frontoparietal regions, nor the full voxel selection performed above chance (all p>0.30). Bars represent group-average accuracy (±sem across subjects), circles represent individual subjects' performance. The star symbols indicate group-level significance: * for p<0.05.

## Discussion

We found that we could take advantage of the expressive power of deep generative neural networks (in particular, VAEs coupled with GANs) to provide a better image space for linear brain decoding. Compared to PCA, which operates in pixel space, our approach produced qualitatively and quantitatively superior results. In particular, we could reliably distinguish the fMRI pattern evoked by one face from another, or determine each face's gender, an outcome which had so far proved elusive[5-9]. We could even decode faces that were not seen but imagined—a true "mind-reading" accomplishment.

One explanation for our method's performance could be that the topology of the VAE-GAN latent space is ideally suited for brain decoding. We already know that this space supports linear operations on faces and facial features[12] (Figure 2). We also know that, by construction (due to the variational training objective of the VAE, and the generative objective of the GAN), nearby points in this space map onto similar-looking but always visually plausible faces. This latent space therefore makes the brain decoding more robust to small mapping errors, partly accounting for our model's performance. In addition to these technical considerations, however, it might simply be that the VAE-GAN latent space is topologically similar to the space of face representations in the human brain. Both types of neural networks (the artificial and the biological ones) are likely to share comparable properties, implicitly reflected in their objective functions: they must somehow "unfold" the complexity of the representation space for face images (in other words, flatten the "face manifold"[15]), making it linear or as close to linear as possible, so that it can be easily manipulated. Although there is unlikely to exist a single solution to this difficult optimization problem (and in fact, there might even be an infinite number of solutions), it is conceivable that all functioning solutions might share common topological features[30]. This speculation that human brain representations are homologous to the latent space of deep generative neural networks could easily be tested in the future, for example using Representational Similarity Analysis[31]. It must be clarified, however, that we do not wish to imply that our particular VAE-GAN implementation is unique in its suitability for brain decoding, or in its resemblance with brain representations; rather, we believe that a whole class of deep generative models could entail similar properties.

Given the explosion of deep generative models in machine learning and computer vision over the last few years[32,33], the successful application of these methods to brain decoding seemed only a matter of time. In fact, several approaches comparable to our own (yet with important differences) have been developed concurrently, and distributed in preprint archives or in conference proceedings over the last year or so. Some used a GAN (without an associated auto-encoder) to produce natural image reconstructions, and trained a brain decoder to associate fMRI response patterns to the GAN latent space[34]. Others did exploit the latent space of an auto-encoder (variational or not), but without the GAN component[35,36]. Yet others attempted to train a GAN to produce natural image reconstructions directly from the brain responses[37,38], rather than using a latent space pre-trained on natural images, and only learning the mapping from brain responses to latent space, as done here. All these pioneering studies produced striking brain-decoded reconstructions of natural scenes or geometric shapes[34-38].

Perhaps most comparable to our own method is the one proposed by Güçlütürk et al[39] to reconstruct face images. They applied GAN training over the CelebA dataset to the output of a convolutional encoder, a standard ConvNet called VGG-Face[40] followed by PCA to reduce its dimensionality to 699 dimensions; then, they learned to map brain responses onto this PCA "latent space" by Bayesian probabilistic inference (maximum a posteriori estimation), and used the GAN to convert the estimated latent decoded vectors into face reconstructions. The test face reconstructions obtained by Güçlütürk et al[39] were already remarkable, even though they used a lower image resolution (64x64 pixels) compared to our own image reconstructions (128x128 pixels). The authors estimated reconstruction accuracy using a structural similarity measure[41], which produced 46.7% similarity for their model (versus about 37% for a baseline PCA-based model). In our case, the structural similarity between original test images and our brain-decoded reconstructions reached 50.5% (range across subjects: [48.4%-52.8%]), while our version of the PCA-based model remained significantly below, around 45.8% (range: [43.5%-47.9%]; $\chi^2(1)=4$, $p<0.05$). Although part of these improvements could be attributed to the increased pixel resolution of our reconstructions, it is clear that our model performs at least as well as the one concurrently developed by Güçlütürk et al[39]. This is particularly important, as our brain decoding method was kept voluntarily much simpler: we used a direct linear mapping between brain responses and latent vectors, rather than the maximum a posteriori probabilistic inference[39]. In our view, the burden of absorbing the complexity of human face representations should lie in the generation of the latent space, rather than in the brain decoder; an effective space should be topologically similar to human brain representations, and thus afford simple (linear) brain decoding. The present results therefore reinforce our hypothesis that state-of-the-art generative models[12,42,43], at least in the realm of face processing, can bring us closer and closer to an adequate model of latent human brain representations.

The proposed brain decoding model holds vast potential for future explorations of face processing and representation in the human brain. As described earlier, this model could be applied to visualize the facial feature selectivity of any voxel or ROI in the brain—directly revealed as an actual face image. The approach could also serve to investigate the brain representation and perception of behaviorally and socially important facial features, such as gender, race, emotion or age; or to study the brain implementation of face-specific attention, memory or mental imagery. One important conclusion of our own explorations, for example, is that occipital voxels greatly contribute to the decoding of perceived faces (Figure 5), but not of imagined faces (Figure 7). Temporal voxels, on the other hand, appear to contribute to both types of trials to a similar extent. This finding may have implications for the understanding of mental imagery and top-down perceptual mechanisms. To help ensure the maximum realization of these promises, we are making the entire fMRI datasets, the brain decoding models for each of the four subjects, and the deep generative neural network used for face encoding and reconstruction fully available to the community (see details in Supplementary Materials).

## Methods

### VAE architecture and GAN training

We trained a "variational auto-encoder" (VAE) deep network (13 layers) using an unsupervised "generative adversarial network" procedure (GAN) for 15 epochs on a labeled database of 202,599 celebrity faces (CelebA dataset[18]). Details of the network architecture are provided in Supplementary Table 1, and particulars of the training procedure can be found in[12]. During GAN training, 3 sub-networks learn complementary tasks (Figure 1A). The Encoder network learns to map a face image onto a 1024-dimensional latent representation (red in Figure 1), which the Generator network can use to produce a novel face image; the Encoder's learning objective is to make the output face image as close as possible to the original image (this reconstruction objective is measured as the L2 loss in the feature space of the Discriminator network, as described in[12]). The Generator network learns to convert latent 1024-D vectors from the latent space into plausible face images. The Discriminator network (6 layers, only used during the training phase) learns to produce a binary decision for each given image (either from the original dataset, or from the Generator output): is the image real or fake? The Discriminator and Generator have opposite objective functions and are updated in alternate steps: the Discriminator is rewarded if it can reliably determine which images come from the Generator (fake) rather than from the dataset (real); the Generator is rewarded if it can produce images that the Discriminator network will not correctly classify. At the end of training, the Discriminator network was discarded, and the Encoder/Generator networks were used as a standard (variational) auto-encoder. Specifically, we used the Encoder to produce 1024-D latent codes for each input face image shown to our human subjects, and these codes served as the design matrix for the fMRI GLM (General Linear Model) analysis (see "Brain decoding" section below). We used the Generator to reconstruct face images based on the output of our "brain decoding" system (a 1024-D latent vector estimate).

### PCA model

Principal Component Analysis (PCA) was used as a baseline (linear) model for face decomposition and reconstruction, as described in Cowen et al[13]. Retaining only the first 1024 principal components (PCs), each image could be turned into a 1024-D code to train our brain decoding system (as detailed below), and output codes could be turned back into face images for visualization using the inverse PCA transform. This baseline model and some of its properties are illustrated in Supplementary Figure S1.

### fMRI scanning procedure

Four subjects (male, 24 to 44 years old) were included in the study, which was performed in accordance with national ethical regulations (Comité de Protection des Personnes, ID RCB 2015-A01801-48). Functional MRI data were collected on a 3T Philips ACHIEVA scanner (gradient echo pulse sequence, TR = 2 s, TE = 10 ms, 41 slices with a 32 channel head coil, slice thickness = 3 mm with 0.2 mm gap, in-plane voxel dimensions 3 x 3 mm). The slices were positioned to cover the entire temporal and occipital lobes. High-resolution anatomical images were also acquired per subject (1×1×1mm voxels, TR = 8.13 ms, TE = 3.74 ms, 170 sagittal slices).

Each subject was tested in 8 scan sessions. Subjects performed between 10 and 14 face runs in each scan session. Each face run started and ended with a 6 s blank interval. Subjects were presented with 88 face stimuli. Each face was presented for 1s, followed by an inter-stimulus interval of 2s (i.e., the inter-trial interval was 3s). The faces subtended 8 degrees of visual angle, and were presented at the center of the screen. Ten test faces (five male and five female) were randomly interspersed among the 88 face stimuli on each run. On alternate runs a different group of 10 test faces was presented (i.e., 20 test faces per subject). Thirty null "fixation" trials were interspersed in each run during which, instead of the face stimulus, a fixation cross was presented on the screen. The face images presented to the subjects in the scanner had been passed once through the VAE-GAN auto-encoder—this was done to ensure that the recorded brain responses concentrated on face or background image properties that could be reliably extracted and reconstructed by the deep generative network. The training image set for each subject was drawn at random from the CelebA dataset, with equal numbers of male and female faces for each run, and disjoint training sets across subjects. A distinct pool of 1,000 potential test faces for each subject was drawn initially at random; we then manually selected from this pool 10 male and 10 female faces, with diverse ages, skin colors, poses and emotions. Again, the 20 test faces were distinct across subjects. To keep subjects alert and encourage them to pay attention to the face stimuli, they were instructed to perform a "1-back" comparison task: press a button as fast as possible whenever the face image was identical to the immediately preceding face. In addition to the 88 face trials, there were 8 one-back trials in each run, and the repeated images were discarded from the brain decoder training procedure (described below). Additionally, whenever the sequence of face images was replaced by a large static gray square (lasting 12s) in the middle of the screen, subjects mentally imagined one specific face image that they had previously chosen among a set of 20 possible faces. For a given subject, only one face image was chosen and studied at length (outside the scanner, between scanning sessions 4 and 5), and then imagined repeatedly throughout scanning sessions 5-8. In odd (respectively even) scanning runs, a unique 12s imagery trial was introduced at the beginning (respectively, the end) of the run. Over the four experimental subjects, the number of recorded imagery trials ranged from 51 to 55 (mean 52). A 6 s blank period followed every imagery trial.

**fMRI Analysis**

fMRI data were processed with SPM 12 (https://www.fil.ion.ucl.ac.uk/spm/software/spm12/). For each participant data from each scan session were slice-time corrected and realigned separately. Then each session was co-registered to the T1 scan from the second MRI session. The data were not normalized or smoothed. The onset and durations of each trial (fixation, training-face, test-face, one-back, or imagery) were entered into a general linear model (GLM) as regressors. Optionally, the 1024 latent vectors (either from the VAE-GAN or the PCA model) of the training face images could be modeled as parametric regressors. Motion parameters were entered as nuisance regressors. The entire design matrix was convolved with SPM's canonical hemodynamic response function (HRF) before the GLM parameters were estimated.

**Brain decoding**

We trained a simple brain decoder (linear regression) to associate the 1024-D latent representation of face images (obtained by running the image through the "Encoder", as described in Figure 1, or using a PCA transform as described above and in Supplementary Figure S1) with the corresponding brain response pattern, recorded when a human subject viewed the same faces in the scanner. This procedure is illustrated in Figure 2A. Each subject saw more than 8,000 faces on average (one presentation each) in a rapid event-related design, and we used the VAE-GAN latent dimensions (or the image projection onto the first 1024 PCs) as 1024 parametric regressors for the BOLD signal (see fMRI analysis section above). These parametric regressors could be positive or negative (since the VAE-GAN latent variables are approximately normally distributed, according to the VAE training objective). An additional categorical regressor ('face vs. fixation' contrast) was added as a constant "bias" term to the model. We verified that the design matrix was "full-rank", i.e. all regressors were linearly independent. This property was expected, because VAE-GAN (and PCA) latent variables tend to be uncorrelated. The linear regression performed by the SPM GLM analysis thus produced a weight matrix W (1025 by $n_{voxels}$ dimensions, where $n_{voxels}$ is the number of voxels in the brain region-of-interest) optimized to predict brain patterns in response to the training face stimuli.

In mathematical terms, we assumed that there exists a linear mapping W between the 1025-dimensional face latent vectors X (including the bias term) and the corresponding brain activation vectors Y (of length $n_{voxels}$), such that:

$$Y = X.W \quad \quad \quad \text{(Eq. 1)}$$

Training the brain decoder consists in finding the optimal mapping W by solving for W:

$$X^T Y = X^T X.W$$

$$W = (X^T X)^{-1}.X^T Y \quad \quad \quad \text{(Eq. 2)}$$

where $X^T X$ is the covariance matrix (1025 by 1025 dimensions) of the latent vectors used for training.

To use this brain decoder in the "testing phase", we simply inverted the linear system, as illustrated in Figure 2B. We presented 20 novel test faces to the same subjects, which had not been seen in the training phase. Each test face was presented on average 52.8 times (range across subjects: [45.4-55.8], randomly interleaved with the training face images) to increase signal-to-noise ratio. The resulting brain activity patterns were simply multiplied by the transposed weight matrix $W^T$ ($n_{voxels}$ by 1025 dimensions) and its inverse covariance matrix to produce an estimate of the 1024 latent face dimensions (in addition to an estimate of the bias term, which was not used further). We then used the Generator network (as illustrated in Figure 1A) to translate the predicted latent vector into a reconstructed face image. For the baseline PCA model, the same logic was applied, but the face reconstruction was obtained via inverse PCA of the decoded 1024-D vector.

Mathematically, testing the brain decoder involves retrieving the latent vector X for each new brain activation pattern Y using the learned weights W. Starting again from Eq. 1, we now solve for X:

$$Y W^T = X.W W^T$$

$$X = Y W^T.(W W^T)^{-1} \quad \quad \quad \text{(Eq. 3)}$$

**Perceptual ratings**

Human judgments for comparing the image quality of the VAE-GAN and PCA face reconstructions were obtained via Amazon Mechanical Turk (AMT) against financial compensation. Each of the 20 test images from the 4 subjects was shown under the word "original", followed by the VAE-GAN and PCA-based reconstructions under the words "option A" and "option B" (with balanced A/B assignment across observers). The instruction stated "Which of the two modified faces is most like the original? Select A or B". Each pair of images was compared 15 times overall, by at least 10 distinct AMT "workers", with each response assignment (VAE-GAN/PCA for option A/B) viewed by at least 5 workers. The experiment thus resulted in a total of 1,200 (=4*20*15) comparisons between the two face reconstruction models.

**Statistics**

Brain decoding accuracy was compared to chance in two ways. The "full recognition" test was successful if and only if the brain-estimated latent vector was closer (as measured by a Pearson correlation) to the target image latent vector than to all 19 distractor image latent vectors. The p-value for each subject was derived from a binomial test with parameters: probability=1/20, number of draws=20. The "pairwise recognition" test involved comparing the brain-estimated latent vector to its target image latent vector and a randomly chosen distractor image latent vector; recognition was successful whenever the brain-estimated latent vector was closer (Pearson correlation) to the target than the distractor vector. As the successive tests for a given target are not independent, a binomial test would not be appropriate here (and would tend to overestimate significance). Instead, we used a non-parametric Monte-Carlo test: according to the null hypothesis, among the 20 (Pearson) correlations of brain-estimated latent vector with test image latent vectors, the rank of the target vector is equally likely to take any value between 1 and 20 (a binomial test would instead assume an intermediate rank to be more likely). We performed $10^6$ random uniform draws of 20 ranks between 1 and 20, and used these draws to compute a surrogate distribution of pairwise decoding performance values under the null hypothesis. For each subject, the p-value was the (upper) percentile of the decoding performance within this distribution. (We verified that, as expected, this produced more conservative significance values than a binomial test with parameters: probability=1/2, number of draws=20*19).

For both the "full" and "pairwise" recognition measures, we compared VAE-GAN and PCA model performance at the group-level with a Friedman non-parametric test. A Friedman test, followed by appropriate post-hoc comparisons, was also used to contrast the three anatomical voxel selections, separately for each decoding model (VAE-GAN or PCA).

The perceptual comparison measure (proportion of VAE-GAN choices) was contrasted against the null hypothesis (equal likelihood of choosing VAE-GAN and PCA reconstructions) using a binomial test with parameters: probability=1/2, number of draws=4*20*15 (four fMRI subjects, 20 test images each, rated 15 times each).

Gender decoding performance, individually and at the group-level, was compared to chance (50%) using a binomial test with parameters: probability=1/2, number of draws for individual tests=20, for group-level tests=4*20 (four subjects, 20 test images each). A Friedman test, followed by appropriate post-hoc comparisons, was used to contrast gender decoding performance across the three anatomical voxel selections.

Imagery decoding performance was measured in a pairwise manner as explained above ("pairwise recognition"): the brain-estimated latent vector was Pearson-correlated with the ground-truth latent vector and the 19 distractor latent vectors; decoding accuracy was the proportion of distractor correlations that were lower than the ground-truth correlation. This performance was averaged across subjects, and compared to chance (50%) using the same Monte-Carlo non-parametric test as above: this time, all $20^4$=160,000 possible draws could be explicitly considered (4 subjects, each with a rank between 1 and 20) to create the surrogate distribution, against which the group-level performance value was compared. A Friedman test, followed by appropriate post-hoc comparisons, was used to contrast imagery decoding performance across the three anatomical voxel selections.

## Acknowledgments

This work was funded by an ERC Consolidator Grant P-CYCLES number 614244 to RV, a BIAL Foundation grant to LR, and an ANR Grant AI-REPS to LR and RV, as well as NVIDIA Corporation through the donation of a Titan V GPU. The authors would like to thank Dr. Isabelle Berry for subject screening, B. Cottereau for assistance with ethics approval, and the staff of the Imaging Center, INSERM-UPS UMR 825 MRI Platform.

## Author contributions

Conceptualization: RV and LR; Neural Network Design and Training: RV; Experimental Investigation: LR; Analysis—fMRI: LR; Analysis—Brain decoding: RV; Writing—Original Draft: RV; Writing—Review and Editing: LR and RV; Funding acquisition: LR and RV.

# Supplementary Materials

**Data availability**

The full fMRI datasets for all four subjects (source data: raw nifti files, event files and stimulus set) are available on OpenNeuro, an open data sharing and analysis platform (https://openneuro.org/datasets/ds001761). The repository also contains the brain decoding models (SPM processed data, as well as Matlab code for producing latent vector estimates from fMRI data) as derivatives. The pre-trained VAE-GAN network with accompanying Python and TensorFlow source code is fully available on GitHub at: https://github.com/rufinv/VAE-GAN-celebA

| Name | Height | Width | Channels | Kernel | Activation |
|---|---|---|---|---|---|
| *Input* | *(128* | *128* | *3)* | | |
| Encoder | 64 | 64 | 192 | 3x3 | elu |
| | 32 | 32 | 256 | 3x3 | elu |
| | 16 | 16 | 384 | 3x3 | elu |
| | 8 | 8 | 512 | 3x3 | elu |
| | 4 | 4 | 768 | 3x3 | elu |
| | 1 | 1 | 1024 | fc | linear |
| Variational Bayes | 1 | 1 | 1024 | fc | - |
| Generator | 4 | 4 | 1024 | fc | elu |
| | 8 | 8 | 512 | 3x3 | elu |
| | 16 | 16 | 384 | 3x3 | elu |
| | 32 | 32 | 256 | 3x3 | elu |
| | 64 | 64 | 192 | 3x3 | elu |
| | 128 | 128 | 3 | 3x3 | elu |
| Discriminator | 64 | 64 | 64 | 4x4 | elu |
| | 32 | 32 | 64 | 4x4 | elu |
| | 16 | 16 | 64 | 4x4 | elu |
| | 8 | 8 | 64 | 4x4 | elu |
| | 4 | 4 | 64 | 4x4 | elu |
| | 1 | 1 | 1 | fc | sigmoid |
| Batch size | 64 | | | | |
| Loss functions | as in *Larsen et al (2016)* (uses feature differences in Discriminator as the auto-encoder reconstruction loss) | | | | |
| Optimizer | Adam, learning rate = 0.0001 | | | | |

**Supplementary Table 1. Architecture of the VAE-GAN network.** fc: fully connected; elu: exponential linear units

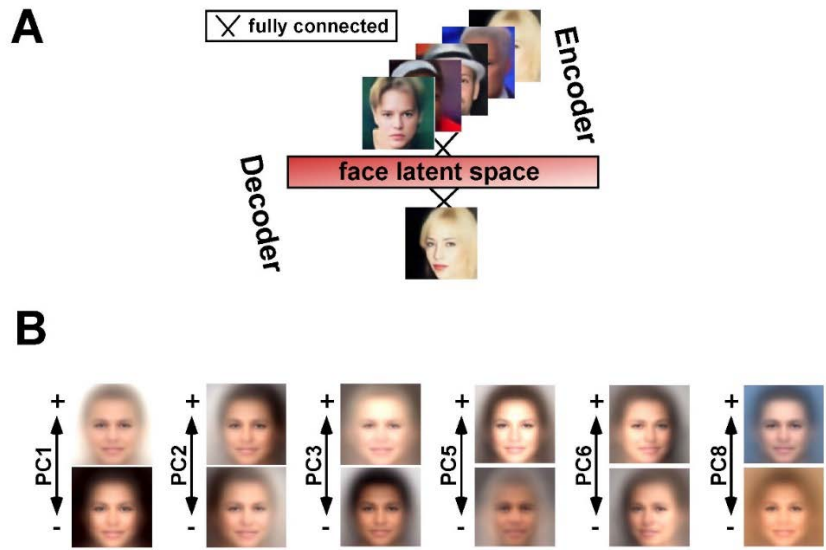

**Figure S1. PCA face decomposition model. A.** Just like in our (VAE-GAN) neural network model, faces are encoded into a latent space of principal components (in red; here, for consistency with the VAE-GAN latent space dimensions, we retained only the first 1024 principal components), and can also be decoded or reconstructed from these components. The main difference, however, is that in PCA the encoding process is a simple linear combination of pixel values. **B.** Some of the first principal components reflect easily interpretable latent dimensions such as face orientation (PC2, PC6), gender (PC5, PC8), skin color (PC3) or background color (PC1, PC8).

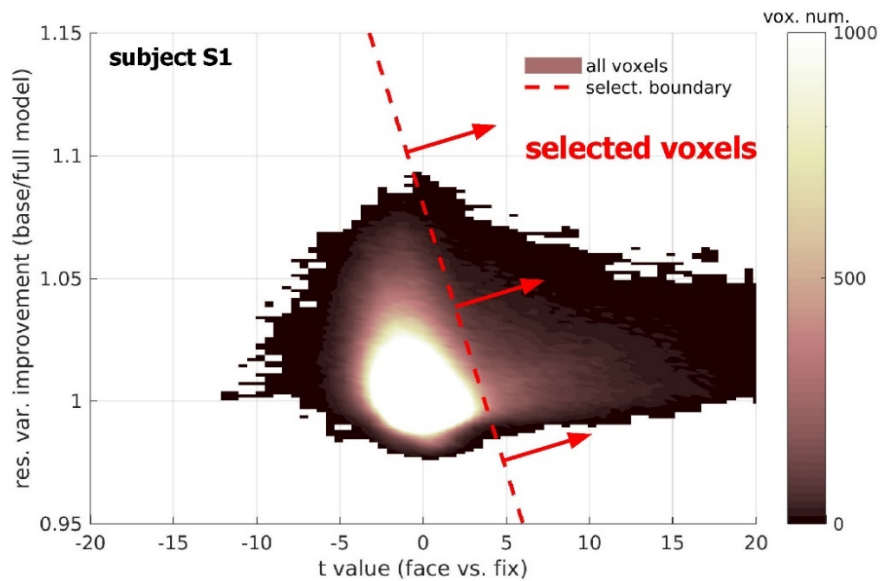

**Figure S2. Voxel selection procedure.** Although fMRI-based face reconstruction was already accurate when using all of the (gray matter) voxels in the brain (about 1.3 million voxels, data not shown here), we found that performance was optimal when selecting a subset of these voxels, combining: (i) a strong visual response to the face stimuli (as measured by a t-test between the "face" condition and the "fixation conditions"; x axis); (ii) an improvement of (adjusted) residual variance when the 1024 latent face dimensions were added as parametric regressors to the baseline GLM (i.e., a GLM with a single binary regressor for face present/absent; y axis). The final selection criterion was a linear combination of these two measures (dashed red boundary). The slope of the boundary was selected, based on the voxel bivariate distribution, so that all voxels with t-values 4 and above (strongly responsive to faces) would be included regardless of their residual variance criterion; and similarly, all voxels with more than 8% adjusted residual variance improvement (strongly sensitive to face latent parameters) would be included, regardless of the t-value. This boundary was chosen based on the fMRI training data of subject S1 (illustrated here), and consequently applied to all other subjects S2-S4, thus limiting the possibility of a spuriously optimal solution. Further, to prevent "double-dipping", the voxel selection was made independently for each subject and each face encoding model (PCA or VAE-GAN), based solely on the BOLD responses collected for training, but not test images.

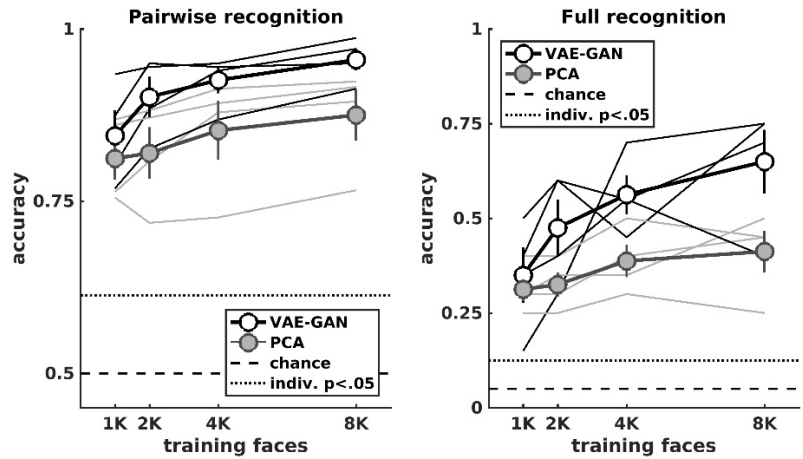

**Figure S3.** Face decoding accuracy (left, pairwise recognition; right, full recognition) as a function of training dataset size. The "full dataset" performance (8 sessions per subject, amounting to ~8K faces) corresponds to the data in Figure 4B-C of the main manuscript. The brain decoding models were also trained with ~1K faces per subject (only the first session), ~2K faces (only the first two sessions) and ~4K faces (only the first four sessions). To facilitate comparison, the same voxel selection (derived using the full dataset) was applied to all training subsets. Circle symbols and thick lines represent mean (±sem) across subjects, thin lines depict individual subject data.

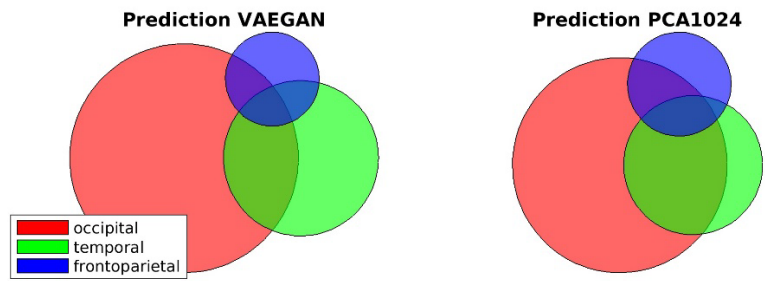

**Figure S4.** Venn diagrams, averaged over subjects, of the proportion of "ground-truth" latent variable variance (for the 20 test images), uniquely or jointly explained by each of the 3 anatomical ROIs (occipital in red, temporal in green, frontoparietal in blue).

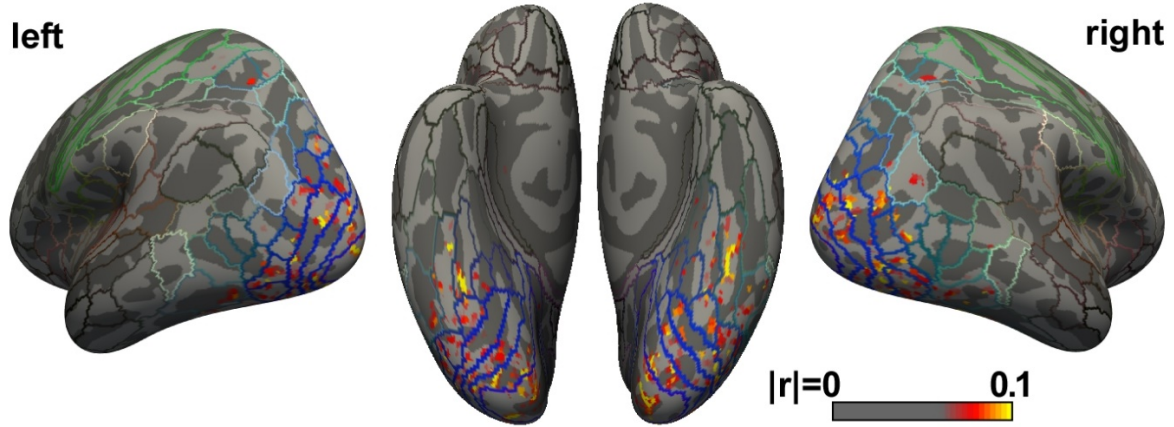

**Figure S5. Mapping of face gender-selective voxels.** A "gender" latent vector was derived by subtracting the average latent description of 10,000 female faces from the average latent description of 10,000 male faces. This vector was then correlated with every column of the "brain decoding" matrix W (see Figure 2): a voxel sensitive to the "gender" property of face images should result in a strongly positive or strongly negative correlation. The subject-averaged absolute value of the correlation r is plotted here on the FreeSurfer average brain. The colored lines indicate the boundaries of standard cortical regions. Gender-selective voxels are found in early visual areas, as well as in regions of the fusiform gyrus.

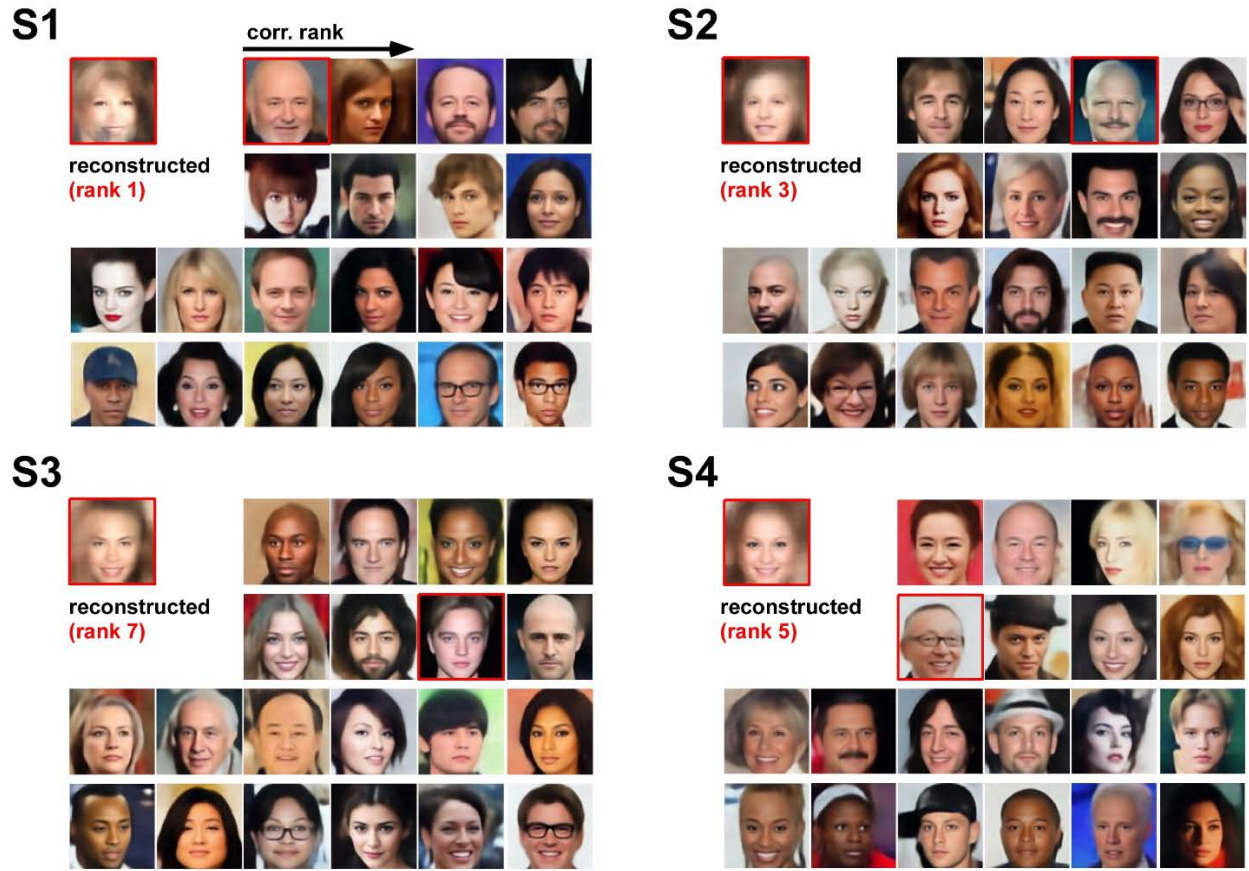

**Figure S6.** Faces reconstructed from mental imagery, based on temporal voxels. For each subject, the reconstruction is shown on the top-left, followed by the 20 candidate faces, ranked by decreasing similarity with the decoded face (correlation of the latent vectors). The image chosen by the subject (i.e., the one that they imagined during the imagery trials) is highlighted in red, and its rank is also reported under the face reconstruction—the lower the rank, the higher the imagery decoding accuracy. Because the brain decoding model was trained during perception (not imagery) conditions, the faces reconstructed from imagery are much less compelling than the ones reconstructed from perception. Nonetheless, one can note interesting details in some reconstructions, such as the "mutton chops" beard for subject S1, or the shape of the eyes in subject S3.